\documentclass[11pt,a4paper,english,superscriptaddress,prd,aps]{revtex4}
\usepackage{amsfonts}
\usepackage{amssymb}
\usepackage{amsmath} 
\usepackage{graphicx}
\usepackage{babel}
\usepackage{hyperref}
\usepackage{color}
\usepackage{slashed}
\usepackage[T1]{fontenc}
\usepackage[latin9]{inputenc}
\usepackage{esint}
\usepackage{subfigure}

\begin{document}

\title{On-shell supersymmetry breaking in the Abelian Chern-Simons-matter model in three dimensions of spacetime}

\author{A. C. Lehum}
\email{lehum@ufpa.br}
\affiliation{\emph{Faculdade de F\'isica, Universidade Federal do Par\'a, 66075-110,
Bel\'em, Par\'a, Brazil}}

\begin{abstract}
This study examines on-shell supersymmetry breaking in the Abelian $\mathcal{N}=1$ Chern-Simons-matter model within a three-dimensional spacetime. The classical Lagrangian is scale invariant, but two-loop radiative corrections to the effective potential break this symmetry, along with gauge and on-shell supersymmetry. To investigate this issue, the renormalization group equation is used to calculate the two-loop effective potential.
\end{abstract}

\maketitle

\section{Introduction} 

Supersymmetric Chern-Simons theories have been an area of study for some time now, with various works \cite{Ivanov:1991fn,Gates:1991qn,Avdeev:1991za,Avdeev:1992jt,ruizruiz:1997jq,Ferrari:2005kx,lehum:2007nf,Ferrari:2010ex,Lehum:2010tt,Gallegos:2011ux} dedicated to understanding them. Recently, these theories have gained significant attention due to their relationship with M2-branes \cite{Gaiotto:2007qi}. The superconformal field theory that describes multiple M2-branes is dual to $D=11$ supergravity on $AdS_{4}\times S^{7}$, and has $\mathcal{N}=8$ supersymmetry. However, the on-shell degrees of freedom of this theory are only bosons and physical fermions, leading to a gauge sector with no on-shell degrees of freedom. These constraints are met by a Chern-Simons-matter theory named BLG theory \cite{Gustavsson:2008dy,Bagger:2007vi,Bagger:2007jr,VanRaamsdonk:2008ft,Bandres:2008ry,Antonyan:2008jf}, which explains the behavior of two M2-branes. By relaxing the prerequisite of manifest $\mathcal{N}=8$ supersymmetry, this approach can be generalized to a $\mathcal{N}=6$ Chern-Simons-matter theory with the gauge group $U_{k}(N)\times U_{-k}(N)$, where $k$ and $-k$ denote the Chern-Simons levels \cite{Aharony:2008ug,Naghdi:2011ex}. It is expected that this theory will attain $\mathcal{N}=8$ supersymmetry when the values of $k$ are either 1 or 2 \cite{Kwon:2009ar,Gustavsson:2009pm,Benna:2009xd}. The quantization of such a model has been extensively investigated in several studies \cite{Faizal:2011en,Faizal:2014dca,Upadhyay:2014oda,Faizal:2012dj,Queiruga:2015fzn,Akerblom:2009gx,Bianchi:2009rf,Bianchi:2009ja,Bianchi:2010cx}. Furthermore, comprehensive computations of the effective superpotential within $\mathcal{N}=2$ superfield theories in three dimensions have been presented in \cite{Buchbinder:2012zd,Buchbinder:2015swa,Lehum:2019msl}.

The one-loop approximation is often sufficient to grasp the key aspects of a field theory. One notable example is the scalar quantum electrodynamics, where Coleman and Weinberg demonstrated that spontaneous gauge symmetry breaking occurs at the one-loop level in four-dimensional spacetime \cite{Coleman:1973jx}. The Coleman-Weinberg mechanism is a valuable tool for understanding how spontaneous symmetry breaking occurs, leading to the generation of masses for Higgs, gauge, and matter fields through radiative corrections in perturbation theory. The method starts with a scale invariant model, and the primary objective is to determine the effective potential for a constant background field and to examine its vacuum properties. Specifically, the goal is to identify conditions under which the effective potential exhibits a minimum. When perturbative corrections to the effective potential produce a nontrivial minimum, accompanied by the spontaneous generation of mass, a dimensional transmutation of one of the model's coupling constants takes place.

The phenomenon of spontaneous symmetry breaking induced by radiative corrections in three-dimensional spacetime is found to occur only after two-loop corrections to the effective potential, as reported in \cite{Tan:1996kz,Tan:1997ew,Dias:2003pw}. It is a well-known fact that perturbative calculations become increasingly complex from the two-loop stage. However, the renormalization group improvement method~\cite{McKeon:1998tr} has proven to be a useful tool to incorporate high-order perturbative effects and has been extensively applied to calculations beyond one-loop approximation \cite{Ahmady:2002qg,Chishtie:2006ck,Elias:2003zm,Chishtie:2005hr,Elias:2004bc,Meissner:2008uw,Dias:2014txa,Quinto:2014zaa,Lehum:2019msl,Souza:2020hjd}. This method works by organizing the effective potential as a power series of logarithms and solving the renormalization group equation (RGE) using perturbative techniques. As a result, insights into the higher-loop effective potential can be gained by utilizing only the one-loop renormalization group functions or, in the case of three-dimensional spacetime, the two-loop renormalization group functions.

The presence of nontrivial $\beta$ and $\gamma$ functions at the two-loop level offers the opportunity to enhance the accuracy of the $V_{eff}$ calculation by utilizing the RGE. The RGE is expressed as
\begin{eqnarray}
\left[\mu\frac{\partial}{\partial\mu}+\beta_{x}\frac{\partial}{\partial x}+\gamma_{\varphi}\varphi\frac{\partial}{\partial\varphi}\right]V_{eff}\left(\varphi;\mu,x,L\right) & =0\,,\label{eq:RGE1}
\end{eqnarray}
\noindent where $x$ denotes the coupling constants of the theory, $\mu$ is the mass scale introduced by the regularization, $\gamma_{\varphi}$ represents the anomalous dimension of scalar field $\varphi$, $L$ is defined as $L =\ln\left(\frac{\varphi^{2}}{\mu}\right)$, and $\phi$ is the vacuum expectation value of $\varphi$. The use of the RGE has been applied to calculate the improved effective potential for a nonsupersymmetric Chern-Simons-matter model in \cite{Dias:2010it}. The results showed significant changes in the model's vacuum properties, highlighting the importance of considering the RGE in a proper analysis of the phase structure. The objective of this study is to employ the RGE to investigate whether the supersymmetric version of the Chern-Simons-matter model also presents a rich phase structure, including spontaneous gauge and supersymmetry breaking.

The structure of this paper is as follows: Section \ref{sec:SCSM-classical} introduces the model and describes how a background field can lead to spontaneous symmetry breaking. It also identifies the relevant physical degrees of freedom. Section \ref{sec:veff2loop} presents the calculation of the two-loop effective action using the RGE method, and shows that the conditions for minimizing the effective potential are consistent with both spontaneous gauge and on-shell supersymmetry breaking. Finally, in Sec. \ref{sec:FR}, we provide concluding remarks.

\section{\label{sec:SCSM-classical}Supersymmetric Chern-Simons-matter model}

In the present section, the classical action of the supersymmetric Chern-Simons-matter model in the ${\cal{N}}=1$ superspace is defined, which allows expressing the action in terms of the physical fields, where the on-shell supersymmetry is realized. The classical action in the $\mathcal{N}=1$ superspace is given by
\begin{eqnarray}
\mathcal{S}  =  \int d^{3}xd^2\theta\left\{ -\frac{1}{4}\Gamma^{\alpha}D^{\beta}D_{\alpha}\Gamma_{\beta}-\frac{1}{2}\overline{\nabla^{\alpha}\Phi}\nabla_{\alpha}\Phi+\frac{\lambda}{4}\left(\overline{\Phi}\Phi\right)^{2}
+\mathcal{L}_{GF}+\mathcal{L}_{FP}
\right\},\label{eq:M1}
\end{eqnarray}
\noindent where the $\Gamma_{\beta}=\Gamma_{\beta}(x,\theta)$ is the Chern-Simons superfield coupled to the massless complex scalar superfield $\Phi=\Phi(x,\theta)$, with a quartic self-interaction. $\nabla^{\alpha}=\left(D^{\alpha}-ie\,\Gamma^{\alpha}\right)$ is the gauge supercovariant derivative. $\mathcal{L}_{GF}$ and $\mathcal{L}_{FP}$ are the gauge fixing and Faddeev-Popov Lagrangians, respectively. The author uses natural units ($\hbar=c=1$), $(-,+,+)$ as the spacetime signature, and the notations and conventions for three-dimensional supersymmetry as found in \cite{Gates:1983nr}.

Alternatively, the action can be written in terms of the component fields. These fields can be defined by the $\theta$ projections given by
\begin{eqnarray}
&&\Gamma_{\alpha}\Big{|}_{\theta=0}=\chi_{\alpha},\hspace{1cm} \frac{1}{2}D^{\alpha}\Gamma_{\alpha}\Big{|}_{\theta=0}=B,\hspace{1cm} D_{\alpha}\Gamma^{\beta}\Big{|}_{\theta=0}=\frac{i}{\sqrt{2}}{(\gamma_\mu)_\alpha}^\beta A^{\mu}-\delta^\beta _\alpha B,\nonumber\\
&& \frac{1}{2}D^\beta D_\alpha\Gamma_\beta\Big{|}_{\theta=0}=\lambda_\alpha,\hspace{1cm}D^{2}\Gamma_{\alpha}\Big{|}_{\theta=0}=\lambda_{\alpha}-i{\partial_\alpha}^{\beta}\chi_\beta~, \nonumber\\
&&\Phi\Big{|}_{\theta=0}=\phi, \hspace{1cm} D_\alpha\Phi\Big{|}_{\theta=0}=\psi_\alpha, \hspace{1cm} D^2\Phi\Big{|}_{\theta=0}=F. 
\label{eq:M1a}
\end{eqnarray}

By integrating over the Grassmann variables and applying the Wess-Zumino gauge ($\chi_\alpha=B=0$), the following expression is obtained:
\begin{eqnarray}
\mathcal{S} & = & \int d^{3}x\Big\{ \frac{1}{2} \epsilon^{\mu\nu\rho}A_{\mu}\partial_{\nu}A_{\rho}-\frac{1}{4}\lambda^\alpha\lambda_\alpha+\bar\psi^{\beta}{(\gamma^\mu)_\beta}^\alpha(\partial_\mu-ieA_\mu)\psi_\alpha+F^*F+\phi^*\Box\phi\nonumber\\
&&-e^2\phi^*\phi A^\mu A_\mu+ieA^\mu\left(\phi^*\partial_\mu\phi-\phi\partial_\mu\phi^*\right)
-\frac{ie}{2}\left(\phi^*\lambda^\alpha\psi_\alpha-\bar\psi^\alpha\lambda_\alpha\phi\right)\nonumber\\
&&+\frac{\lambda}{2}\phi^*\phi\bar\psi^\alpha\psi_\alpha
+\frac{\lambda}{4}\left(\bar\psi^\alpha\bar\psi_\alpha\phi^2+\psi^\alpha\psi_\alpha{\phi^*}^2\right)+\frac{\lambda}{2}\left(F^*\phi+\phi^*F\right)\phi^*\phi+\mathcal{L}_{GF}+\mathcal{L}_{FP}\Big\}.\label{eq:M1b}
\end{eqnarray}

The focus of the study is on the Coleman-Weinberg mechanism. To calculate the effective potential, a shift in the fields $\phi$, $\phi^*$, $F$ and $F^*$ is considered, given by
\begin{subequations}
\begin{eqnarray}
\phi \rightarrow\phi+\frac{\varphi}{\sqrt{2}}=\frac{1}{\sqrt{2}}\left(\phi_1+i\phi_2\right)+\frac{\varphi}{\sqrt{2}}, \label{eq:M2a}\\
\phi^* \rightarrow\phi^*+\frac{\varphi}{\sqrt{2}}=\frac{1}{\sqrt{2}}\left(\phi_1-i\phi_2\right)+\frac{\varphi}{\sqrt{2}}, \label{eq:M2b}\\
F \rightarrow F+\frac{\mathfrak{f}}{\sqrt{2}}=\frac{1}{\sqrt{2}}\left(f_1+if_2\right)+\frac{\mathfrak{f}}{\sqrt{2}}, \label{eq:M2c}\\
F^* \rightarrow F^*+\frac{\mathfrak{f}}{\sqrt{2}}=\frac{1}{\sqrt{2}}\left(f_1-if_2\right)+\frac{\mathfrak{f}}{\sqrt{2}},\label{eq:M2d}
\end{eqnarray}
\end{subequations}

\noindent where $\varphi$ and $\mathfrak{f}$ are real background (constant) fields.

In terms of the classical background fields, the action \eqref{eq:M1b} can be cast as
\begin{eqnarray}
\mathcal{S} & = & \int d^{3}x\Big\{ \frac{1}{2} \epsilon^{\mu\nu\rho}A_{\mu}\partial_{\nu}A_{\rho}-\frac{1}{4}\lambda^\alpha\lambda_\alpha+\bar\psi^{\beta}{(\gamma^\mu)_\beta}^\alpha(\partial_\mu-ieA_\mu)\psi_\alpha\nonumber\\
&&-\frac{1}{2}\left(f_1^2+f_2^2+2\mathfrak{f}f_1+\mathfrak{f}^2\right)+\frac{1}{2}\phi_1\Box\phi_1+\frac{1}{2}\phi_2\Box\phi_2\nonumber\\
&&-\frac{e^2}{2}\left(\phi_1^2+\phi_2^2+2\varphi\phi_1+\varphi^2 \right) A^\mu A_\mu 
-\frac{e}{2} A^\mu\left(\phi_1\partial_\mu\phi_2-\phi_2\partial_\mu\phi_1\right)\nonumber\\
&&-\frac{ie}{2}\left[\frac{(\phi_1-i\phi_2)}{2}\lambda^\alpha\psi_\alpha-\bar\psi^\alpha\lambda_\alpha\frac{(\phi_1+i\phi_2)}{2}\right]\nonumber\\
&&+\frac{\lambda}{4}\left(\phi_1^2+\phi_2^2+2\varphi\phi_1+\varphi^2 \right)\bar\psi^\alpha\psi_\alpha\nonumber\\
&&+\frac{\lambda}{8}\bar\psi^\alpha\bar\psi_\alpha\left[\phi_1^2-\phi_2^2+2\varphi(\phi_1+i\phi_2)+2i\phi_1\phi_2+\varphi^2\right]\nonumber\\
&&+\frac{\lambda}{8}\psi^\alpha\psi_\alpha\left[\phi_1^2-\phi_2^2+2\varphi(\phi_1-i\phi_2)-2i\phi_1\phi_2+\varphi^2 \right]\nonumber\\
&&+\frac{\lambda}{4}\left[(f_1+\mathfrak{f}) (\varphi +\phi_1)+f_2\phi_2\right] \left(\phi_1^2+\phi_2^2+\varphi^2+2\varphi\phi_1\right) \nonumber\\
&&+\frac{ie\varphi}{2} {\lambda^\alpha\frac{(\bar\psi_\alpha-\psi_\alpha)}{\sqrt{2}}}
-e \varphi A^\mu \partial_\mu \phi_2 
+\mathcal{L}_{GF}+\mathcal{L}_{FP} \Big\}.\label{eq:M1c}
\end{eqnarray}

In order to eliminate the mixing between $A^\mu$ and $\phi_2$ and between $\lambda^\alpha$ and $\frac{(\bar\psi_\alpha-\psi_\alpha)}{\sqrt{2}}$ in the last line of \eqref{eq:M1c}, a supersymmetric $R_{\xi}$ gauge fixing is used, $\mathcal{F}_{G}=\left(D^{\alpha}\Gamma_{\alpha}+i \frac{\xi}{2} e\varphi\frac{(\bar\Phi-\Phi)}{\sqrt{2}}\right)$. This is achieved by including the gauge fixing and Faddeev-Popov actions,
\begin{eqnarray}\label{ceq6a} 
S_{GF+FP}&=&\int{d^5z}\Big\{\frac{1}{2\xi} \left(D^{\alpha}\Gamma_{\alpha}+i\frac{\xi}{2}e\varphi\frac{(\bar\Phi-\Phi)}{\sqrt{2}}\right)^2 +\bar{C}D^2C+\frac{\xi e^2\varphi^2}{4}\bar{C}C+\frac{\xi e^2\varphi}{4}\bar{C}\frac{(\bar\Phi+\Phi)}{\sqrt{2}} C\Big\}\nonumber\\
&=&\int{d^3x}\Big\{ \frac{1}{2\xi} \lambda^\alpha\lambda_\alpha
-\frac{\xi}{8}e^2\varphi^2\frac{(\bar\psi^\alpha-\psi^\alpha)}{\sqrt{2}}\frac{(\bar\psi_\alpha-\psi_\alpha)}{\sqrt{2}} 
+\frac{\xi}{4}e^2\varphi^2\phi_2 f_{2}\nonumber\\
&&+e\varphi A^\mu\partial_\mu\phi_2
-i\frac{e\varphi}{2}\lambda^{\alpha}\frac{(\bar\psi_\alpha-\psi_\alpha)}{\sqrt{2}}+\mathcal{L}_{ghosts}
\Big\},
\end{eqnarray}
\noindent where $C$ and $\bar{C}$ are the ghost superfields and $\mathcal{L}_{ghosts}$ is the Lagrangian involving the ghost fields. 

Integrating out the auxiliary fields $f_1$ and $f_2$, we can express the action in terms of the physical fields as follows,
\begin{eqnarray}
\mathcal{S} & = & \int d^{3}x\Big\{ \frac{1}{2} \epsilon^{\mu\nu\rho}A_{\mu}\partial_{\nu}A_{\rho}-\frac{e^2\varphi^2}{2}A^{\mu}A_{\mu}-\frac{1}{4}\left(1-\frac{2}{\xi}\right)\lambda^{\alpha}\lambda_{\alpha}\nonumber\\
&&+\frac{1}{2}\left(|\partial_\mu \phi_1|^2-\frac{15\lambda^2\varphi^4}{16}\phi_1^2\right)+\frac{1}{2}\left[|\partial_\mu \phi_2|^2-\frac{(3\lambda+\xi\lambda e^4)\varphi^4}{16}\phi_2^2\right]\nonumber\\
&&+i\bar\psi^\alpha{(\gamma^\mu)_\alpha}^\beta \partial_{\mu}\psi_\beta+\frac{\lambda\varphi^2}{4}\bar\psi^\alpha\psi_\alpha+\frac{\lambda\varphi^2}{8}(\bar\psi^\alpha\bar\psi_{\alpha}+\psi^{\alpha}\psi_\alpha)\nonumber\\
&&-\frac{\xi}{8}e^2\varphi^2\frac{(\bar\psi^\alpha-\psi^\alpha)}{\sqrt{2}}\frac{(\bar\psi_\alpha-\psi_\alpha)}{\sqrt{2}}+\mathrm{interactions}\Big\}.\label{eq:M1d}
\end{eqnarray}
\noindent One interesting feature of the calculation is the complete absence of the classical field $\mathfrak{f}$ after the integration over the auxiliary fields $f_1$ and $f_2$.

The imaginary part $\phi_2$ of $\phi$ becomes the Goldstone boson due to spontaneous gauge symmetry breaking. However, it should be noted that $\phi_2$ becomes a nonphysical field after introducing the typical $R_\xi$ gauge fixing. This results in a gauge dependent mass term for $\phi_2$, and the degree of freedom corresponding to $\phi_2$ is absorbed by the Chern-Simons field $A^{\mu}$ in the process of becoming massive.

The fermion mass matrix can be written as
\begin{eqnarray}
\mathcal{L}_m &=& \frac{\lambda\varphi^2}{8}\left(\bar\psi^\alpha,\psi^\alpha\right)\left(\begin{array}{cc}
1 & 1\\
 1 & 1  
\end{array}\right)\left(\begin{array}{c}
\bar\psi_\alpha \\
\psi_\alpha  
\end{array}\right)
-\frac{\xi}{8}e^2\varphi^2\frac{(\bar\psi^\alpha-\psi^\alpha)}{\sqrt{2}}\frac{(\bar\psi_\alpha-\psi_\alpha)}{\sqrt{2}}\nonumber\\
&=&\frac{m_f}{2}\frac{(\bar\psi^\alpha+\psi^\alpha)}{\sqrt{2}}\frac{(\bar\psi_\alpha+\psi_\alpha)}{\sqrt{2}}+\frac{\tilde{m}_f}{2}\frac{i(\bar\psi^\alpha-\psi^\alpha)}{\sqrt{2}}\frac{i(\bar\psi_\alpha-\psi_\alpha)}{\sqrt{2}}\nonumber\\
&=&\frac{m_f}{2}\psi_+^\alpha ~{\psi_+}_\alpha+\frac{\tilde{m}_f}{2}\psi^\alpha_- ~{\psi_-}_\alpha.
\end{eqnarray}

\noindent After diagonalizing the fermion mass matrix, two states are obtained: one massive state with mass $m_f=\frac{\lambda \varphi^2}{2}$, represented by $\psi_+^{\alpha}=\frac{1}{\sqrt{2}}(\bar\psi^\alpha+\psi^\alpha)$, and another state with mass $\tilde{m}_f=\frac{\xi}{4} e^2\varphi^2$ represented by $\psi_{-}^{\alpha}=\frac{i}{\sqrt{2}}(\bar\psi^\alpha-\psi^\alpha)$, which is interpreted as the Goldstino. Similar to the case of $\phi_2$, the fermionic state $\psi_{-}$ will also be a nonphysical field due to the introduction of the $R_\xi$ gauge fixing, and it will have a gauge dependent mass term.

If spontaneous symmetry breaking occurs at some level, i.e., if the background field $\varphi$ assumes a nonvanishing value $v$ at the minimum of the effective potential, the Chern-Simons field acquires a mass term proportional to $m_a=e^2v^2$, while the real part $\phi_1$ of the complex field $\phi$ acquires a squared mass $m_b^2=(\frac{15\lambda^2v^4}{16}+\text{radiative corrections})$. The fermionic matter state $\psi^\alpha_+$ acquires a squared mass $m_f^2=\frac{\lambda^2v^4}{4}$ due to the tree-level Higgs mechanism, which is different from its supersymmetric partner $\phi$.

In the case where the Coleman-Weinberg mechanism takes place, the squared mass ratio between bosonic and fermionic matter fields is given by
\begin{equation} \frac{m_b^2}{m_f^2}=\frac{15}{4}+(\text{radiative corrections}),\label{eq:mb/mf} \end{equation}
which indicates the spontaneous on-shell supersymmetry breaking if $v\neq 0$.

\section{\label{sec:veff2loop}Determination of the effective potential}

Before studying the two-loop effective potential for the Chern-Simons-matter model, it is useful to discuss some aspects of the potential at the classical level. For simplicity, consider an action for a real scalar superfield $\Phi(z)=(\phi(x)+\theta^\alpha \psi_\alpha(x) -\theta^2 F(x))$ given by
\begin{eqnarray}
S &=& \int{d^5z}\left[ \frac{1}{2}\Phi D^2\Phi -\frac{\lambda}{4}\Phi^4 \right]=\int{d^3x}\left[\frac{1}{2}\phi\Box\phi+\frac{1}{2}F^2+\lambda F\phi^3 \right],
\end{eqnarray}
\noindent where the terms involving the fermionic field $\psi$ have been suppressed.

\begin{figure}[h] \begin{centering} \includegraphics[width=14cm]{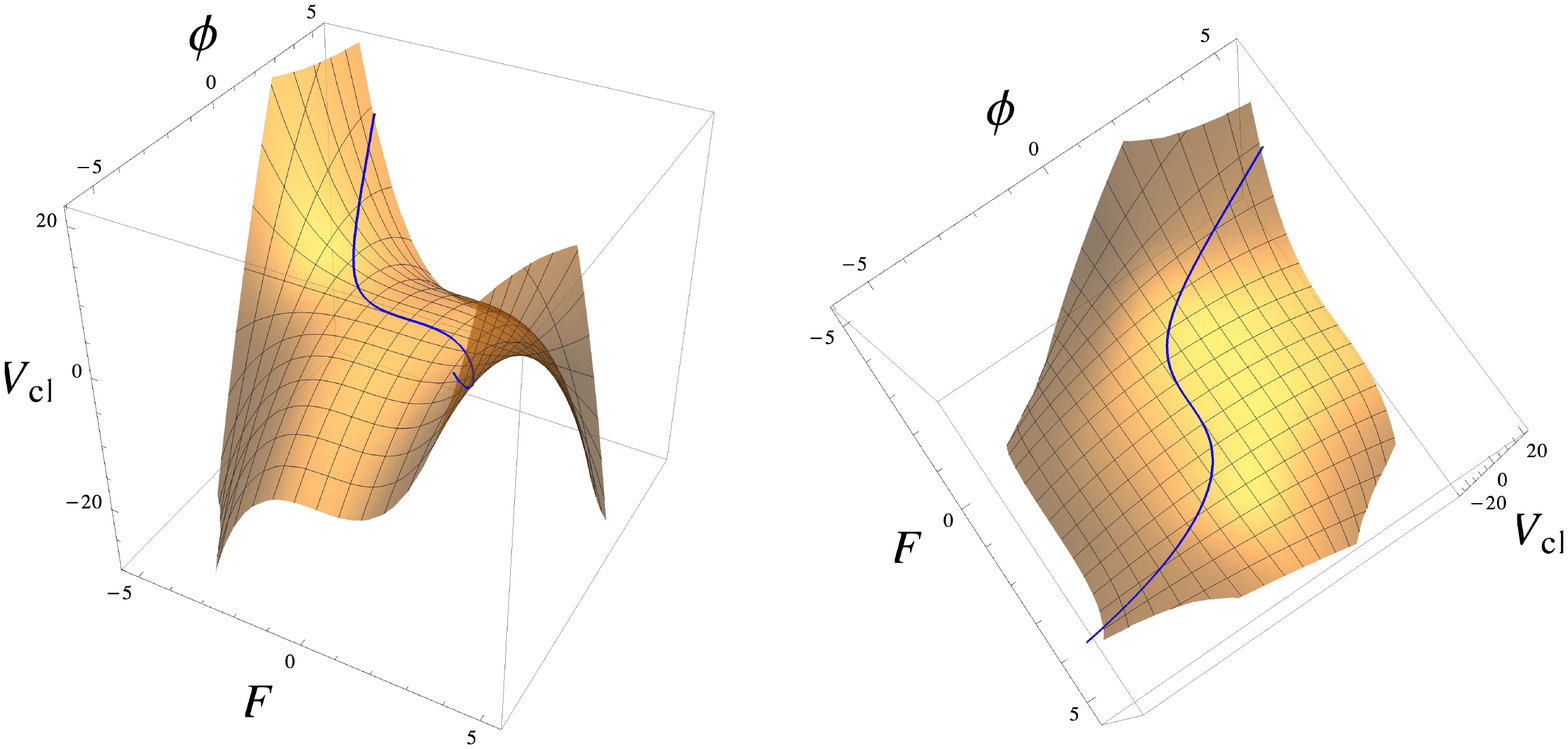} \par\end{centering}\caption{\label{fig_V_c} Plot of the classical potential in the presence of the auxiliary field $F$. The line in blue corresponds to the classical path $F=-\lambda\phi^3$.} \end{figure}

The above action leads to a classical potential given by 
\begin{eqnarray}\label{vclassicalF}
V_{cl}=-\frac{F^2}{2}-\lambda F\phi^3,
\end{eqnarray} 
which is depicted in Fig. \ref{fig_V_c}.

\begin{figure}[h] \begin{centering} \includegraphics[width=7cm]{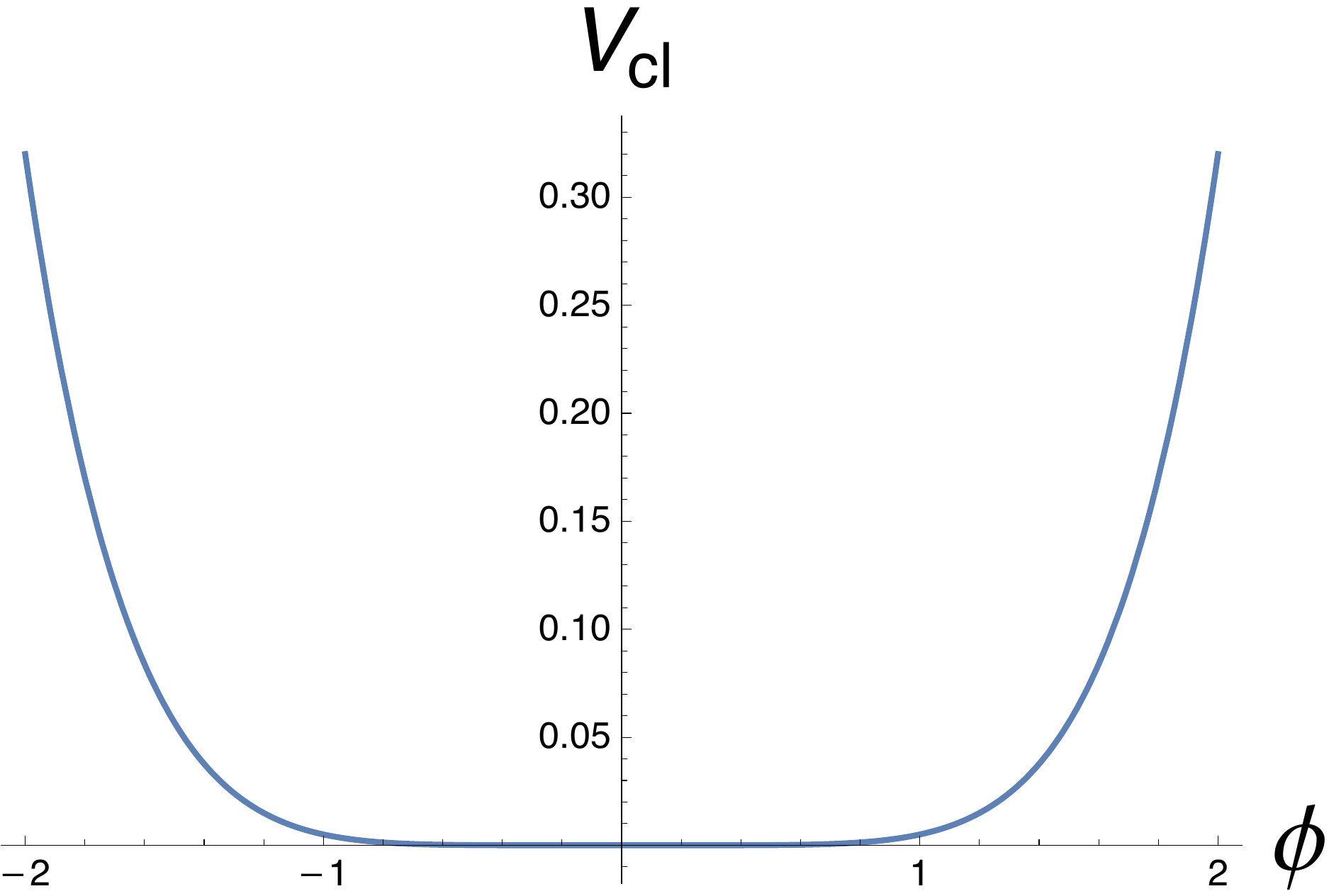} \par\end{centering}\caption{\label{fig_V_c2} Plot of the classical potential after eliminating the auxiliary field $F$.} \end{figure}

Some attempts have been made in the literature to compute the effective potential while keeping the intermediate steps of the calculation manifestly supersymmetric, without eliminating the auxiliary field $F$ from the beginning~\cite{Burgess:1983nu,Miller:1983fe,Miller:1983ri,Gallegos:2011ag,Gallegos:2011ux,Maluf:2012ie}. However, it has been observed that the potential \eqref{vclassicalF} is not bounded from below when expressed in terms of the auxiliary field $F$. Therefore, to compute radiative corrections to the classical potential, it is advisable to eliminate the auxiliary field $F$ by employing its equation of motion $F=-\lambda \phi^3$ from the beginning, as suggested by Murphy and O'Raifeartaigh \cite{Murphy:1983ag}. This choice leads to a well-defined classical potential, depicted in Fig. \ref{fig_V_c2}, given by 
\begin{eqnarray}
V_{cl}=\frac{\lambda^2}{2}\phi^6.
\end{eqnarray}    

Thus, in order to compute the radiative corrections to the classical potential of the Chern-Simons-matter model, we will follow the approach proposed by Murphy and O'Raifeartaigh and eliminate the auxiliary field $F$ using its equation of motion from the beginning. Afterwards, we will compute the radiative corrections to the classical potential in order to obtain the two-loop effective potential of the model. 

The effective potential $V_{eff}\left(\varphi\right)$ can be obtained using the RGE with the beta function and anomalous dimension from the literature~\cite{Avdeev:1991za}. The two-loop beta functions and anomalous dimension for the Chern-Simons-matter model are expressed in terms of the redefined gauge coupling constant $y=e^{2}$, and can be written as
\begin{subequations}\label{eq:RGEfunctions}
\begin{align}
\beta_{\lambda} & =\frac{1}{64\pi^2}\left(9\lambda^{3}+\lambda^{2}y-6\lambda y^{2}-4y^{3}\right)\thinspace,\\
\beta_{y} & =0\thinspace,\\
\gamma_{\varphi} & =\frac{1}{64\pi^2}\left(\frac{3\lambda^{2}}{4}-\frac{5y^{2}}{4}\right)+\frac{1}{64\pi^2}\left(\xi_1 y^2+\xi_2 \lambda y\right)\thinspace,
\end{align}
\end{subequations}
\noindent where $\xi_1$ and $\xi_2$ are gauge dependent calculable coefficients. 

The two-loop beta functions and anomalous dimension for the Chern-Simons-matter model have been computed in Ref.~\cite{Avdeev:1991za} using the Landau gauge. To address a possible gauge dependence issue of the effective potential \cite{Jackiw:1974cv,Nielsen:1975fs}, we have included the possible gauge dependent part of the anomalous dimension. However, the actual values of $\xi_1$ and $\xi_2$ are not important, as the two-loop effective potential is completely gauge invariant, which is a unique feature of the two-loop approximation (in three dimensions). In higher-order corrections, gauge dependent quantities, such as daisies, can appear, as discussed in \cite{Bazeia:1988pz,deLima:1989yf,Andreassen:2014eha}, but this is not the case in the approximation presented here.

We shall use the following ansatz for $V_{eff}\left(\varphi\right)$
\begin{equation}\label{ansatz1}
V_{eff}\left(\varphi\right)=\frac{1}{32}\varphi^{6}S\left(L\right)\thinspace,
\end{equation}
where
\begin{equation}
S\left(L\right)=A\left(y,\lambda\right)+B\left(y,\lambda\right)L+C\left(y,\lambda\right)L^{2}+\cdots\thinspace,\label{eq:KeffAnsatz2}
\end{equation}
with $A,\thinspace B,\thinspace C,\thinspace\ldots$ defined as power series of the coupling constants $y$ and $\lambda$, and $L$ is defined by
\begin{eqnarray}\label{eq:defL}
L =\ln\left(\frac{\varphi^{2}}{\mu}\right).
\end{eqnarray}

In the adopted shorthand notation, any of the two couplings in the model is denoted by the symbol $x$. For instance, $y^{n}\lambda^{m}$ is represented as $x^{m+n}$. It can be observed that the two-loop beta function and anomalous dimension are of order ${\cal O}(x^3)$ and ${\cal O}(x^2)$, respectively.

In the comparison with the action\,(\ref{eq:M1c}), we can identify $A\left(y,\lambda\right)$ as
\begin{equation}
A\left(y,\lambda\right)=\lambda^2+{\cal O}\left(x^{3}\right)\thinspace.\label{eq:A}
\end{equation}
Actually, the value of $A\left(y,\lambda\right)$ will be fixed by the Coleman-Weinberg normalization of the effective potential,
\begin{equation}
\frac{1}{6!}\frac{d^{6}V_{eff}\left(\varphi\right)}{d^{6}\varphi}\Big{|}_{\varphi=\sqrt{\mu}}=\frac{\lambda^2}{32}\thinspace,\label{eq:CWcondition}
\end{equation}
\noindent where we have chosen $\varphi=\sqrt{\mu}$ as the renormalization scale.

In order to compute the effective potential through RGE given in \eqref{eq:RGE1}, it is necessary to use the fact that from
\eqref{eq:defL} it follows that $\partial_{L}=\frac{1}{2}\varphi\partial_{\varphi}=-\mu\partial_{\mu}$. By using \eqref{ansatz1} to compute \eqref{eq:RGE1} an alternative form for the RGE can be found, which is
\begin{equation}
\left[-\left(1+2\gamma_{\varphi}\right)\partial_{L}+\beta_{\lambda}\partial_{\lambda}+6\gamma_{\varphi}\right]S\left(L\right)=0\thinspace,\label{eq:RGE2}
\end{equation}
and will be used hereafter. 

By inserting the ansatz given by \eqref{eq:KeffAnsatz2} into \eqref{eq:RGE2}, a series of equations can be obtained by collecting the resulting expression by orders of $L$. The first two of these equations is quoted below:
\begin{equation}
-\left(1+2\gamma_{\varphi}\right)B\left(y,\lambda\right)+\beta_{\lambda}\partial_{\lambda}A\left(y,\lambda\right)+6\gamma_{\varphi}A\left(y,\lambda\right)=0\thinspace,\label{eq:orderL0}
\end{equation}
and
\begin{equation}
-2\left(1+2\gamma_{\varphi}\right)C\left(y,\lambda\right)+\beta_{\lambda}\partial_{\lambda}B\left(y,\lambda\right)+6\gamma_{\varphi}B\left(y,\lambda\right)=0\thinspace.\label{eq:orderL1}
\end{equation}

Given that all functions that appear in these equations are defined as power series of the couplings $x$, the equation \eqref{eq:orderL0} can be written as
\begin{multline}
-\left(B^{\left(2\right)}+B^{\left(3\right)}+B^{\left(4\right)}+\cdots\right)-2\left(\gamma_{\varphi}^{\left(2\right)}+\gamma_{\varphi}^{\left(3\right)}+\cdots\right)\left(B^{\left(2\right)}+B^{\left(3\right)}+B^{\left(4\right)}+\cdots\right)\\
+\left(\beta_{\lambda}^{\left(3\right)}+\beta_{\lambda}^{\left(4\right)}+\cdots\right)\frac{\partial}{\partial\lambda}\left(A^{\left(2\right)}+A^{\left(3\right)}+A^{\left(4\right)}+\cdots\right)\\
+6\left(\gamma_{\varphi}^{\left(2\right)}+\gamma_{\varphi}^{\left(3\right)}+\cdots\right)\left(A^{\left(2\right)}+A^{\left(3\right)}+A^{\left(4\right)}+\cdots\right)=0\thinspace.\label{eq:orderL0powers}
\end{multline}
\noindent where the numbers in the superscripts represent the power of $x$ for each term.

Organizing the above equation in order of $x$, we find
\begin{eqnarray}
&&-B^{\left(2\right)}-B^{\left(3\right)}+\left[-B^{\left(4\right)} -2\gamma_{\varphi}^{\left(2\right)}B^{\left(2\right)}+\beta_{\lambda}^{\left(3\right)}\frac{\partial A^{\left(2\right)}}{\partial\lambda} +6\gamma_{\varphi}^{\left(2\right)}A^{\left(2\right)}\right]+\cdots=0\thinspace,\label{eq:orderL0powers}
\end{eqnarray}
where it is easy to see that $B^{\left(2\right)}=B^{\left(3\right)}=0$. By the use of \eqref{eq:A}, the above equation in ${\cal O}(x^4)$ can be cast as
\begin{eqnarray}
B^{\left(4\right)}&=& 2\lambda\beta_{\lambda}^{\left(3\right)}+6\lambda^2\gamma_{\varphi}^{\left(2\right)}\nonumber\\
&=&\frac{\lambda}{128\pi^2} \left(42 \lambda^3-16 y^3+3 \lambda  (4 \xi_1-13) y^2+4 \lambda ^2 (3 \xi_2+1) y\right)\thinspace.\label{eq:B4}
\end{eqnarray}

The corrections of order $x^4 L$ in $S_{eff}$ can be obtained through a two-loop calculation of the effective potential. However, the coefficients of $\beta_{\lambda}^{(4)}$ and $\gamma_{\varphi}^{(3)}$ are unknown, which would appear from higher loop corrections. Therefore, it is not possible to use  (\ref{eq:orderL0}) to calculate further coefficients of $B$ or $A$. As a result, this equation does not provide information on higher-order loop contributions to $S_{eff}$.

The effective potential can be computed up to order $x^4L$, which corresponds to two-loop order in perturbation theory. The ansatz in \eqref{ansatz1}, combined with \eqref{eq:A} and \eqref{eq:B4}, allows us to express the effective potential as
\begin{eqnarray}
V_{eff}(\varphi)=\frac{\varphi^6}{32}\left[A^{(2)}+B^{(4)}\ln\left(\frac{\varphi^2}{\mu}\right)+\delta_\lambda\right],\end{eqnarray}
\noindent where $\delta_\lambda$ is the counterterm that satisfies the renormalization condition \eqref{eq:CWcondition}.

Imposing the Coleman-Weinberg condition, \eqref{eq:CWcondition}, and choosing the renormalization scale $\varphi=\sqrt{\mu}$, finally we find
\begin{eqnarray}
V_{eff}(\varphi)=\frac{\lambda^2}{32}\varphi^6+\frac{\lambda \left[42 \lambda^3-16 y^3+3 \lambda  (4 \xi_1-13) y^2+4 \lambda^2 (3 \xi_2+1) y \right]}{40960\pi^2} \varphi^6 \left[-49+10 \ln\left(\frac{\varphi^2}{\mu}\right)\right].\label{eq:veffL}
\end{eqnarray}
\noindent As a result, we find that the effective potential is apparently gauge dependent. But, as we will see, the gauge dependence is removed after imposition of the conditions that minimize the effective potential.

The conditions to minimize the effective potential are given by
\begin{subequations}\label{eq:conditions}
\begin{align}
\frac{dV_{eff}\left(\varphi\right)}{d\varphi}\Big{|}_{\varphi=\sqrt{\mu}}=0\thinspace,\label{eq:gap1}\\
\frac{d^2 V_{eff}\left(\varphi\right)}{d\varphi^2}\Big{|}_{\varphi=\sqrt{\mu}}>0\thinspace,\label{eq:mass1}
\end{align}
\end{subequations}
\noindent where, because of arbitrariness of the renormalization scale, we set the scale of renormalization $\varphi=\sqrt{\mu}$ to be the minimum of the effective potential. The extremum condition, \eqref{eq:gap1}, 
\begin{eqnarray}
\frac{\lambda  \mu ^{5/2} \left(6 \lambda  \left(640 \pi ^2-959 \lambda ^2\right)+2192 y^3-411 \lambda  (4 \xi_1-13) y^2-548 \lambda ^2 (3 \xi_2+1) y\right)}{20480 \pi ^2}=
0,
\end{eqnarray}
\noindent yields a nontrivial perturbative solution for $\lambda$, given by
\begin{eqnarray}
\lambda\approx -\frac{137 y^3}{240 \pi^2}+\mathcal{O}(y^6).\label{lambda1}
\end{eqnarray}

Upon substituting the expression for $\lambda$ given in \eqref{lambda1} into \eqref{eq:veffL}, it follows that
\begin{eqnarray}\label{effective_potential}
V_{eff}=\frac{137 y^6 \varphi ^6}{184320 \pi ^4}\left[-1+3 \ln\left(\frac{\varphi^2}{\mu}\right)\right],
\end{eqnarray}
\noindent with
\begin{eqnarray}
&&\frac{d V_{eff}\left(\varphi\right)}{d\varphi}\Big{|}_{\varphi=\sqrt{\mu}}=0;\\
&&m^2_{\varphi}=\frac{d^2 V_{eff}\left(\varphi\right)}{d\varphi^2}\Big{|}_{\varphi=\sqrt{\mu}}=
\frac{15 \lambda^2 \mu^2}{16}+\frac{125 \lambda y^3 \mu^2}{256 \pi^2}
=
\frac{137y^6}{5120 \pi^4}\mu^2>0.
\end{eqnarray}

In the present model, the Coleman-Weinberg mechanism dislocated the classical global minimum $\varphi=0$ into a local maximum, thus relocating the minimum of the effective potential to $\varphi=v=\sqrt{\mu}$ as expected. It is worth noting that the two-loop effective potential \eqref{effective_potential} is invariant under gauge transformations. However, higher-order corrections can generate gauge dependent quantities, such as daisies~\cite{Bazeia:1988pz,deLima:1989yf,Andreassen:2014eha}, and therefore require more thorough analysis.

As discussed previously, the occurrence of the Coleman-Weinberg mechanism results in simultaneous spontaneous gauge and off-shell supersymmetry breaking. This is indeed the case in this model, where the Chern-Simons field becomes massive with $m_a^2=e^4v^4=y^2\mu^2$, and the real part $\phi_1$ of the complex field $\phi$ acquires a squared mass $m_b^2=(\frac{15\lambda^2v^4}{16}+\text{radiative corrections})=\frac{137y^6}{5120 \pi^4}\mu^2$. Additionally, the massive fermionic state $\frac{1}{\sqrt{2}}(\bar\psi^\alpha+\psi^\alpha)$ acquires a squared mass $m_f^2=\frac{\lambda^2 v^4}{4}=\frac{18769 y^6}{230400 \pi ^4}\mu^2$, indicating a spontaneous on-shell supersymmetry breaking.

In regards to supersymmetry breaking, a crucial aspect is the positivity of the potential. It has been noted that the effective potential $V_{eff}$, given in \eqref{effective_potential}, is negative at $\varphi=\sqrt{\mu}$, with $V_{eff}(\mu)=-\frac{137 y^6 \mu^3}{184320 \pi^4}$. This result appears to contradict our understanding of supersymmetry breaking. However, it should be emphasized that the effective potential obtained from the RGE \eqref{eq:RGE1} is the effective potential up to a constant term. This constant of integration was not taken into consideration when the ansatz \eqref{ansatz1} was used. In the Appendix, some arguments are presented to justify that the effective potential evaluated at its minimum is indeed positive, on the order of $V_{min}=e^{c/y^2}\mu^3+\mathcal{O}(y^6)$, where $y$ is perturbative with $y\ll1$, and $c\sim 400$ is a constant.

\section{\label{sec:FR}Final Remarks}

In this work, the on-shell supersymmetry breaking in the Abelian $\mathcal{N}=1$ Chern-Simons-matter model is investigated. The classical Lagrangian is scale invariant, which is broken by two-loop radiative corrections to the effective potential. To conclude this, the author computed the two-loop effective potential by the use of renormalization group equation showing that gauge symmetry and on-shell supersymmetry is spontaneously broken induced by radiative corrections, i.e., through Coleman-Weinberg mechanism. The Coleman-Weinberg mechanism turned the classical global minimum $\varphi=0$ into a local maximum, dislocating the minimum of the effective potential to $\varphi=\sqrt{\mu}$, just as it should be. It was found that the two-loop effective potential is gauge invariant, but it is important to remark that high order corrections can generate gauge dependent quantities such as daisies~\cite{Bazeia:1988pz,deLima:1989yf,Andreassen:2014eha}, and a more careful analysis has to be made.

An important aspect of supersymmetry breaking is the positivity of the effective potential. Even though apparently the value of $V_{eff}$ at the minimum is negative, it was argued that the effective potential obtained from the renormalization group equation \eqref{eq:RGE1} corresponds to the effective potential apart from a constant, where this constant of integration was not taken into account by the use of the ansatz \eqref{ansatz1}. Some argumentations can be done to justify that the value of the effective potential evaluated at its minimum is in fact positive, of order of $V_{min}=e^{c/y^2}\mu^3+\mathcal{O}(y^6)$, with perturbative $y\ll1$ and $c\sim 400$ being a constant. These argumentations are presented in the Appendix.

Finally, an extension of this work for higher order corrections could be instructive to investigate the possible gauge dependence and stability of the radiatively generated minimum.


\section*{ACKNOWLEDGMENTS}

The author is grateful to A. J. da Silva and A. F. Ferrari for useful discussions.

\appendix
\section{POSITIVITY OF THE EFFECTIVE POTENTIAL}

One of the key features of supersymmetric field theory is the positivity of the effective potential. It is widely known that the value of the effective potential at its minimum must be positive in the case of spontaneous supersymmetry breaking \cite{Gates:1983nr}. In this paper, it is computed the two-loop effective potential of the on-shell supersymmetric Chern-Simons-matter theory in a three-dimensional spacetime. However, an unexpected result emerged from the analysis. Specifically, the value of $V_{eff}$ at the minimum, i.e., $\phi=\sqrt{\mu}$, was found to be $V_{eff}(\mu)=-\frac{137 y^6 \mu^3}{184320 \pi^4}<0$. Nevertheless, it is important to emphasize that the effective potential obtained from the RGE \eqref{eq:RGE1} corresponds to the effective potential apart from a constant. This constant of integration was not considered in the ansatz \eqref{ansatz1}.

In this Appendix, argumentations are presented to justify the positivity of the effective potential evaluated at its minimum. The effective potential constant term $V_0$ is assumed to have the form given by
\begin{eqnarray}\label{v0_ansatz}
V_0=f(\lambda,y)\mu^3, 
\end{eqnarray}
\noindent where $V_0$ depends on the coupling constants but not on the background field $\varphi$. The ansatz \eqref{v0_ansatz} is expected to be of order $\mu^3$ by dimensional analysis.

Plugging \eqref{v0_ansatz} into the RGE \eqref{eq:RGE1}, we find
\begin{eqnarray}
\left[\mu\frac{\partial}{\partial\mu}+\beta_{\lambda}\frac{\partial}{\partial \lambda}\right]V_{0}=\mu^3\left[3f(\lambda,y)+\beta_\lambda\frac{df(\lambda,y)}{d\lambda}\right]=0.
\end{eqnarray}

It is possible to obtain a solution for $f(\lambda, y)$ by solving its linear first-order differential equation. One such solution is given by
\begin{eqnarray}
V_0&=&\mu^3~\mathrm{exp}\left( 96\pi^2\frac{\sqrt{11} \ln \left(9 \lambda ^2+4 y^2+10 \lambda  y\right)
-\sqrt{11} \ln (\lambda -y)^2
+28 \tan ^{-1}\left(\frac{9 \lambda +5 y}{\sqrt{11} y}\right)}
{23 \sqrt{11} y^2}\right)\nonumber\\
&\approx &\mu^3~\exp \left(\frac{192 \pi^2 \left(11 \ln (2)+14 \sqrt{11} \tan ^{-1}\left(\frac{5}{\sqrt{11}}\right)\right)}{253 y^2}-\frac{137}{5}\right)\nonumber\\
&\sim & \mu^3~\exp\left({\frac{400}{y^2}}\right)>0,
\end{eqnarray}
\noindent where $\lambda$ was expended around its value in the minimum of $V_{eff}$, as given by Eq.\eqref{lambda1}, and considered a perturbative regime where $y\ll1$.

By adding $V_0$ to the effective potential \eqref{effective_potential}, the minimum of the effective potential can be determined. It can be cast as
\begin{eqnarray}
V_{min}=V_0+V_{eff}\Big{|}_{\varphi=\sqrt{\mu}}=\mu^3 \exp\left({\frac{400}{y^2}}\right)-\frac{137 y^6 \mu^3}{184320 \pi^4}>0,
\end{eqnarray}
\noindent that is evident that $V_{min}$ is positive, which guarantees the positivity of the effective potential.

\newpage 


\begin{thebibliography}{10}

\bibitem{Ivanov:1991fn}
E.~A.~Ivanov,
\textit{Chern-Simons matter systems with manifest N=2 supersymmetry},
Phys. Lett. B \textbf{268}, 203-208 (1991)
doi:10.1016/0370-2693(91)90804-Y

\bibitem{Gates:1991qn}
S.~J.~Gates, Jr. and H.~Nishino,
\textit{Remarks on the N=2 supersymmetric Chern-Simons theories},
Phys. Lett. B \textbf{281}, 72-80 (1992)
doi:10.1016/0370-2693(92)90277-B

\bibitem{Avdeev:1991za}
L.~V.~Avdeev, G.~V.~Grigorev and D.~I.~Kazakov,
\textit{Renormalizations in Abelian Chern-Simons field theories with matter},
Nucl. Phys. B \textbf{382}, 561-580 (1992)
doi:10.1016/0550-3213(92)90659-Y

\bibitem{Avdeev:1992jt}
L.~V.~Avdeev, D.~I.~Kazakov and I.~N.~Kondrashuk,
\textit{Renormalizations in supersymmetric and nonsupersymmetric nonAbelian Chern-Simons field theories with matter},
Nucl. Phys. B \textbf{391}, 333-357 (1993)
doi:10.1016/0550-3213(93)90151-E

\bibitem{ruizruiz:1997jq}
F.~Ruiz Ruiz and P.~van Nieuwenhuizen,
\textit{Supersymmetric Yang-Mills Chern-Simons theory},
Nucl. Phys. B Proc. Suppl. \textbf{56}, 269-274 (1997)
doi:10.1016/S0920-5632(97)00335-6
[arXiv:hep-th/9701052 [hep-th]].

\bibitem{Ferrari:2005kx}
A.~F.~Ferrari, M.~Gomes, A.~Y.~Petrov and A.~J.~da Silva,
\textit{Supersymmetric non-Abelian noncommutative Chern-Simons theory},
Phys. Lett. B \textbf{638}, 275-282 (2006)
doi:10.1016/j.physletb.2006.05.031
[arXiv:hep-th/0511059 [hep-th]].

\bibitem{lehum:2007nf}
A.~C.~Lehum, A.~F.~Ferrari, M.~Gomes and A.~J.~da Silva,
\textit{Spontaneous gauge symmetry breaking in a SUSY Chern-Simons model},
Phys. Rev. D \textbf{76}, 105021 (2007)
doi:10.1103/PhysRevD.76.105021
[arXiv:0709.3280 [hep-th]].

\bibitem{Ferrari:2010ex}
A.~F.~Ferrari, E.~A.~Gallegos, M.~Gomes, A.~C.~Lehum, J.~R.~Nascimento, A.~Y.~Petrov and A.~J.~da Silva,
\textit{Coleman-Weinberg mechanism in a three-dimensional supersymmetric Chern-Simons-Matter model},
Phys. Rev. D \textbf{82}, 025002 (2010)
doi:10.1103/PhysRevD.82.025002
[arXiv:1004.0982 [hep-th]].

\bibitem{Lehum:2010tt}
A.~C.~Lehum and A.~J.~da Silva,
\textit{Spontaneous breaking of superconformal invariance in (2+1)D supersymmetric Chern-Simons-matter theories in the large N limit},
Phys. Lett. B \textbf{693}, 393-398 (2010)
doi:10.1016/j.physletb.2010.08.059
[arXiv:1008.1173 [hep-th]].

\bibitem{Gallegos:2011ux}
E.~A.~Gallegos and A.~J.~da Silva,
\textit{Dynamical (super)symmetry vacuum properties of the supersymmetric Chern-Simons-matter model},
Phys. Rev. D \textbf{85}, 125012 (2012)
doi:10.1103/PhysRevD.85.125012
[arXiv:1111.2886 [hep-th]].

\bibitem{Gaiotto:2007qi}
D.~Gaiotto and X.~Yin,
\textit{Notes on superconformal Chern-Simons-Matter theories},
JHEP \textbf{08}, 056 (2007)
doi:10.1088/1126-6708/2007/08/056
[arXiv:0704.3740 [hep-th]].

\bibitem{Gustavsson:2008dy}
A.~Gustavsson,
\textit{Selfdual strings and loop space Nahm equations},
JHEP \textbf{04}, 083 (2008)
doi:10.1088/1126-6708/2008/04/083
[arXiv:0802.3456 [hep-th]].

\bibitem{Bagger:2007vi}
J.~Bagger and N.~Lambert,
\textit{Comments on multiple M2-branes},
JHEP \textbf{02}, 105 (2008)
doi:10.1088/1126-6708/2008/02/105
[arXiv:0712.3738 [hep-th]].

\bibitem{Bagger:2007jr}
J.~Bagger and N.~Lambert,
\textit{Gauge symmetry and supersymmetry of multiple M2-branes},
Phys. Rev. D \textbf{77}, 065008 (2008)
doi:10.1103/PhysRevD.77.065008
[arXiv:0711.0955 [hep-th]].

\bibitem{VanRaamsdonk:2008ft}
M.~Van Raamsdonk,
\textit{Comments on the Bagger-Lambert theory and multiple M2-branes},
JHEP \textbf{05}, 105 (2008)
doi:10.1088/1126-6708/2008/05/105
[arXiv:0803.3803 [hep-th]].

\bibitem{Bandres:2008ry}
M.~A.~Bandres, A.~E.~Lipstein and J.~H.~Schwarz,
\textit{Studies of the ABJM Theory in a Formulation with Manifest SU(4) R-Symmetry},
JHEP \textbf{09}, 027 (2008)
doi:10.1088/1126-6708/2008/09/027
[arXiv:0807.0880 [hep-th]].

\bibitem{Antonyan:2008jf}
E.~Antonyan and A.~A.~Tseytlin,
\textit{On 3d N=8 Lorentzian BLG theory as a scaling limit of 3d superconformal N=6 ABJM theory},
Phys. Rev. D \textbf{79}, 046002 (2009)
doi:10.1103/PhysRevD.79.046002
[arXiv:0811.1540 [hep-th]].

\bibitem{Aharony:2008ug}
O.~Aharony, O.~Bergman, D.~L.~Jafferis and J.~Maldacena,
\textit{N=6 superconformal Chern-Simons-matter theories, M2-branes and their gravity duals},
JHEP \textbf{10}, 091 (2008)
doi:10.1088/1126-6708/2008/10/091
[arXiv:0806.1218 [hep-th]].

\bibitem{Naghdi:2011ex}
M.~Naghdi,
\textit{A Monopole Instanton-Like Effect in the ABJM Model},
Int. J. Mod. Phys. A \textbf{26}, 3259-3273 (2011)
doi:10.1142/S0217751X11053833
[arXiv:1106.0907 [hep-th]].

\bibitem{Kwon:2009ar}
O.~K.~Kwon, P.~Oh and J.~Sohn,
\textit{Notes on Supersymmetry Enhancement of ABJM Theory},
JHEP \textbf{08}, 093 (2009)
doi:10.1088/1126-6708/2009/08/093
[arXiv:0906.4333 [hep-th]].

\bibitem{Gustavsson:2009pm}
A.~Gustavsson and S.~J.~Rey,
\textit{Enhanced N=8 Supersymmetry of ABJM Theory on R**8 and R**8/Z(2)},
[arXiv:0906.3568 [hep-th]].

\bibitem{Benna:2009xd}
M.~K.~Benna, I.~R.~Klebanov and T.~Klose,
\textit{Charges of Monopole Operators in Chern-Simons Yang-Mills Theory},
JHEP \textbf{01}, 110 (2010)
doi:10.1007/JHEP01(2010)110
[arXiv:0906.3008 [hep-th]].

\bibitem{Faizal:2011en}
M.~Faizal,
\textit{M-Theory on Deformed Superspace},
Phys. Rev. D \textbf{84}, 106011 (2011)
doi:10.1103/PhysRevD.84.106011
[arXiv:1111.0213 [hep-th]].

\bibitem{Faizal:2012dj}
M.~Faizal,
\textit{$M$-Theory in the Gaugeon Formalism},
Commun. Theor. Phys. \textbf{57}, 637-640 (2012)
doi:10.1088/0253-6102/57/4/20
[arXiv:1201.1220 [hep-th]].

\bibitem{Faizal:2014dca}
M.~Faizal and S.~Upadhyay,
\textit{Spontaneous Breaking of the BRST Symmetry in the ABJM theory},
Phys. Lett. B \textbf{736}, 288-292 (2014)
doi:10.1016/j.physletb.2014.07.040
[arXiv:1407.6188 [hep-th]].

\bibitem{Upadhyay:2014oda}
S.~Upadhyay and D.~Das,
\textit{ABJM theory in Batalin\textendash{}Vilkovisky formulation},
Phys. Lett. B \textbf{733}, 63-68 (2014)
doi:10.1016/j.physletb.2014.04.019
[arXiv:1404.2633 [hep-th]].

\bibitem{Queiruga:2015fzn}
J.~M.~Queiruga, A.~C.~Lehum and M.~Faizal,
\textit{K\"ahlerian effective potentials for Chern\textendash{}Simons-matter theories},
Nucl. Phys. B \textbf{902}, 58-68 (2016)
doi:10.1016/j.nuclphysb.2015.11.007
[arXiv:1511.03586 [hep-th]].

\bibitem{Akerblom:2009gx}
N.~Akerblom, C.~Saemann and M.~Wolf,
\textit{Marginal Deformations and 3-Algebra Structures},
Nucl. Phys. B \textbf{826}, 456-489 (2010)
doi:10.1016/j.nuclphysb.2009.08.012
[arXiv:0906.1705 [hep-th]].

\bibitem{Bianchi:2009rf}
M.~S.~Bianchi, S.~Penati and M.~Siani,
\textit{Infrared Stability of N = 2 Chern-Simons Matter Theories},
JHEP \textbf{05}, 106 (2010)
doi:10.1007/JHEP05(2010)106
[arXiv:0912.4282 [hep-th]].

\bibitem{Bianchi:2009ja}
M.~S.~Bianchi, S.~Penati and M.~Siani,
\textit{Infrared stability of ABJ-like theories},
JHEP \textbf{01}, 080 (2010)
doi:10.1007/JHEP01(2010)080
[arXiv:0910.5200 [hep-th]].

\bibitem{Bianchi:2010cx}
M.~S.~Bianchi and S.~Penati,
\textit{The Conformal Manifold of Chern-Simons Matter Theories},
JHEP \textbf{01}, 047 (2011)
doi:10.1007/JHEP01(2011)047
[arXiv:1009.6223 [hep-th]].

\bibitem{Buchbinder:2012zd}
I.~L.~Buchbinder, B.~S.~Merzlikin and I.~B.~Samsonov,
\textit{Two-loop effective potentials in general N=2, d=3 chiral superfield model},
Nucl. Phys. B \textbf{860}, 87-114 (2012)
doi:10.1016/j.nuclphysb.2012.02.013
[arXiv:1201.5579 [hep-th]].

\bibitem{Buchbinder:2015swa}
I.~L.~Buchbinder and B.~S.~Merzlikin,
\textit{On effective K\"ahler potential in N=2 , d=3 SQED},
Nucl. Phys. B \textbf{900}, 80-103 (2015)
doi:10.1016/j.nuclphysb.2015.09.002
[arXiv:1505.07679 [hep-th]].

\bibitem{Lehum:2019msl}
A.~C.~Lehum, J.~R.~Nascimento, A.~Yu.~Petrov and H.~Souza,
\textit{Renormalization Group Improvement of the Superpotential for the N=2 Chern-Simons-matter model},
Phys. Rev. D \textbf{101}, no.4, 045005 (2020)
doi:10.1103/PhysRevD.101.045005
[arXiv:1911.10846 [hep-th]].

\bibitem{Coleman:1973jx}
S.~R.~Coleman and E.~J.~Weinberg,
\textit{Radiative Corrections as the Origin of Spontaneous Symmetry Breaking}, 
Phys. Rev. D \textbf{7}, 1888-1910 (1973)
doi:10.1103/PhysRevD.7.1888

\bibitem{Tan:1996kz}
P.~N.~Tan, B.~Tekin and Y.~Hosotani,
\textit{Spontaneous symmetry breaking at two loop in 3-D massless scalar electrodynamics},
Phys. Lett. B \textbf{388}, 611-620 (1996)
doi:10.1016/S0370-2693(96)01191-4
[arXiv:hep-th/9607233 [hep-th]].

\bibitem{Tan:1997ew}
P.~N.~Tan, B.~Tekin and Y.~Hosotani,
\textit{Maxwell Chern-Simons scalar electrodynamics at two loop},
Nucl. Phys. B \textbf{502}, 483-515 (1997)
doi:10.1016/S0550-3213(97)00495-1
[arXiv:hep-th/9703121 [hep-th]].

\bibitem{Dias:2003pw}
A.~G.~Dias, M.~Gomes and A.~J.~da Silva,
\textit{Dynamical breakdown of symmetry in (2+1) dimensional model containing the Chern-Simons field},
Phys. Rev. D \textbf{69}, 065011 (2004)
doi:10.1103/PhysRevD.69.065011
[arXiv:hep-th/0305043 [hep-th]].
 
\bibitem{McKeon:1998tr}
D.~G.~C.~McKeon,
\textit{Summing logarithms in quantum field theory: The renormalization group},
Int. J. Theor. Phys. \textbf{37}, 817-826 (1998)
doi:10.1023/A:1026620630263

\bibitem{Ahmady:2002qg}
M.~Ahmady, V.~Elias, D.~McKeon, A.~Squires and T.~G.~Steele,
\textit{Renormalization group improvement of effective actions beyond summation of leading logarithms},
Nucl. Phys. B \textbf{655}, 221-249 (2003)
doi:10.1016/S0550-3213(03)00008-7
[arXiv:hep-ph/0211227 [hep-ph]].

\bibitem{Elias:2003zm}
V.~Elias, R.~B.~Mann, D.~G.~C.~McKeon and T.~G.~Steele,
\textit{Radiative electroweak symmetry breaking revisited},
Phys. Rev. Lett. \textbf{91}, 251601 (2003)
doi:10.1103/PhysRevLett.91.251601
[arXiv:hep-ph/0304153 [hep-ph]].

\bibitem{Elias:2004bc}
V.~Elias, R.~B.~Mann, D.~G.~C.~McKeon and T.~G.~Steele,
\textit{Higher order stability of a radiatively induced 200-GeV Higgs mass},
Phys. Rev. D \textbf{72}, 037902 (2005)
doi:10.1103/PhysRevD.72.037902
[arXiv:hep-ph/0411161 [hep-ph]].

\bibitem{Chishtie:2005hr}
F.~Chishtie, V.~Elias, R.~B.~Mann, D.~McKeon and T.~Steele,
\textit{Stability of subsequent-to-leading-logarithm corrections to the effective potential for radiative electroweak symmetry breaking},
Nucl. Phys. B \textbf{743}, 104-132 (2006)
doi:10.1016/j.nuclphysb.2006.03.005
[arXiv:hep-ph/0509122 [hep-ph]].

\bibitem{Chishtie:2006ck}
F.~Chishtie, V.~Elias, R.~B.~Mann, D.~McKeon and T.~Steele,
\textit{On the standard approach to renormalization group improvement},
Int. J. Mod. Phys. E \textbf{16}, 1681-1685 (2007)
doi:10.1142/S0218301307006095
[arXiv:hep-th/0609199 [hep-th]].

\bibitem{Meissner:2008uw}
K.~A.~Meissner and H.~Nicolai,
\textit{Renormalization Group and Effective Potential in Classically Conformal Theories},
Acta Phys. Polon. B \textbf{40}, 2737-2752 (2009)
[arXiv:0809.1338 [hep-th]].

\bibitem{Dias:2014txa}
A. G. Dias, A. F. Ferrari, J. D. Gomez, A. A. Natale, A. G. Quinto,
\textit{Non-perturbative fixed points and renormalization group improved effective potential},
Phys. Lett. B \textbf{739}, 8-12 (2014)
doi:10.1016/j.physletb.2014.10.017
[arXiv:1407.1879 [hep-ph]].

\bibitem{Quinto:2014zaa}
A.~G.~Quinto, A.~F.~Ferrari and A.C.~Lehum,
\textit{Renormalization group improvement and dynamical breaking of symmetry in a supersymmetric Chern-Simons-matter model},
Nucl. Phys. B \textbf{907}, 664-677 (2016)
doi:10.1016/j.nuclphysb.2016.04.015
[arXiv:1405.6118 [hep-th]].

\bibitem{Souza:2020hjd}
H.~Souza, L.~Ibiapina Bevilaqua and A.~C.~Lehum,
\textit{Renormalization group improvement of the effective potential in six dimensions},
Phys. Rev. D \textbf{102}, no.4, 045004 (2020)
doi:10.1103/PhysRevD.102.045004
[arXiv:2005.03973 [hep-th]].

\bibitem{Dias:2010it}
A.~G.~Dias and A.~F.~Ferrari,
\textit{Renormalization Group and Conformal Symmetry Breaking in the Chern-Simons Theory Coupled to Matter},
Phys. Rev. D \textbf{82}, 085006 (2010)
doi:10.1103/PhysRevD.82.085006
[arXiv:1006.5672 [hep-th]].

\bibitem{Gates:1983nr}
S.~J.~Gates, M.~T.~Grisaru, M.~Rocek and W.~Siegel,
\textit{Superspace Or One Thousand and One Lessons in Supersymmetry},
Front. Phys. \textbf{58}, 1-548 (1983)
1983,
ISBN 978-0-8053-3161-5
[arXiv:hep-th/0108200 [hep-th]].

\bibitem{Burgess:1983nu}
C.~P.~Burgess,
\textit{Supersymmetry breaking in three-dimensions},
Nucl. Phys. B \textbf{216}, 459-468 (1983)
doi:10.1016/0550-3213(83)90295-X
  
\bibitem{Miller:1983fe}
R.~D.~C.~Miller,
\textit{A Tadpole Supergraph Method for the Evaluation of {SUSY} Effective Potentials},
Nucl. Phys. B \textbf{228}, 316-332 (1983)
doi:10.1016/0550-3213(83)90327-9

\bibitem{Miller:1983ri}
R.~D.~C.~Miller,
\textit{The Auxiliary Field Tadpole Method of Effective Potential Evaluation: Its Extension to {SUSY} Gauge Theories},
Nucl. Phys. B \textbf{229}, 189-204 (1983)
doi:10.1016/0550-3213(83)90360-7

\bibitem{Gallegos:2011ag}
E.~A.~Gallegos and A.~J.~da Silva,
\textit{Supergraph techniques for D=3,N=1 broken supersymmetric theories},
Phys. Rev. D \textbf{84}, 065009 (2011)
doi:10.1103/PhysRevD.84.065009
[arXiv:1102.1989 [hep-th]].

\bibitem{Maluf:2012ie}
R.~V.~Maluf and A.~J.~da Silva,
\textit{Two-loop effective potential for the Wess-Zumino model in 2 + 1 dimensions},
Phys. Rev. D \textbf{87}, no.4, 045022 (2013)
doi:10.1103/PhysRevD.87.045022
[arXiv:1207.1706 [hep-th]].

\bibitem{Murphy:1983ag}
T.~Murphy and L.~O'Raifeartaigh,
\textit{A note on supersymmetry breaking in (1+1)-dimensions},
Nucl. Phys. B \textbf{218}, 484-492 (1983)
doi:10.1016/0550-3213(83)90376-0

\bibitem{Jackiw:1974cv}
R.~Jackiw,
\textit{Functional evaluation of the effective potential},
Phys. Rev. D \textbf{9}, 1686 (1974)
doi:10.1103/PhysRevD.9.1686

\bibitem{Nielsen:1975fs}
N.~K.~Nielsen,
\textit{On the Gauge Dependence of Spontaneous Symmetry Breaking in Gauge Theories},
Nucl. Phys. B \textbf{101}, 173-188 (1975)
doi:10.1016/0550-3213(75)90301-6

\bibitem{Bazeia:1988pz}
D.~Bazeia,
\textit{Gauge invariance and daisies},
Phys. Lett. B \textbf{207}, 53-55 (1988)
doi:10.1016/0370-2693(88)90885-4

\bibitem{deLima:1989yf}
A.~F.~de Lima and D.~Bazeia,
\textit{Gauge invariance and Nielsen identities},
Z. Phys. C \textbf{45}, 471 (1990)
doi:10.1007/BF01549677

\bibitem{Andreassen:2014eha}
A.~Andreassen, W.~Frost and M.~D.~Schwartz,
\textit{Consistent Use of Effective Potentials},
Phys. Rev. D \textbf{91}, no.1, 016009 (2015)
doi:10.1103/PhysRevD.91.016009
[arXiv:1408.0287 [hep-ph]].

\end{thebibliography}
\end{document}